# Spectroscopy of low–spin states in 157Dy: Search for evidence of enhanced octupole correlations

S. N. T. Majola[1, 2, 3]*, R. A. Bark[3], L. Bianco[12], T. D. Bucher[3], S. P. Bvumbi[2], D. M. Cullen[4, 7], P. E. Garrett[12], P. T. Greenlees[4], D. Hartley[8], J Hirvonen[4], U. Jakobsson[4], P. M. Jones[3], R. Julin[4], S. Juutinen[4], S. Ketelhut[4], B. V. Kheswa[2, 3], A. Korichi[13], E. A. Lawrie[3, 5], P. L. Masiteng[2], B. Maqabuka[3], L. Mdletshe[1, 3], A. Minkova[14], J. Ndayishimye[3], P. Nieminen[4], R. Newman[3], B. M. Nyakó[9], S. S. Ntshangase[1], P. Peura[4], P. Rahkila[4], L. L. Riedinger[10], M. Riley[6], D. Roux[11], P. Ruotsalainen[4], J. Saren[4], J. F. Sharpey-Schafer[5], C. Scholey[4], O. Shirinda[3], A. Sithole[1, 5], J.Sorri[4]†, S. Stolze[4]‡, J. Timár[9], J. Uusitalo[4] and G. Zimba[2].

[1] University of Zululand, Department of Physics and Engineering, kwaDlangezwa, 3886, South Africa
[2] Department of Physics, University of Johannesburg, P.O. Box 524, Auckland Park 2006, South Africa
[3] iThemba LABS, P. O. Box 722, Somerset–West 7129, South Africa
[4] University of Jyvaskyla, Department of Physics, P.O. Box 35, FI-40014 University of Jyväskylä, Finland
[5] University of the Western Cape, Department of Physics, P/B X17, Bellville 7535, South Africa
[6] Department of Physics, Florida State University, Tallahassee, FL 32306, USA
[7] Schuster Laboratory, University of Manchester, Manchester M13 9PL, United Kingdom.
[8] Department of Physics, U.S. Naval Academy, Annapolis, Maryland 21402, USA
[9] MTA Atomki, P.O. Box 51, H–4001 Debrecen, Hungary
[10] University of Tennessee, Department of Physics and Astronomy, Knoxville, Tennessee 37996, USA
[11] Department of Physics, Rhodes University, P.O. Box 74, Grahamstown 6140, South Africa
[12] University of Guelph, Department of Physics, Guelph, Ontario NIG 2WI, Canada
[13] CSNSM–IN2P3–CNRS, F–91405 Orsay Campus, France
[14] University of Sofia, Faculty of Physics, Sofia 1164, Bulgaria
Present addresses: † Sodankylä Geophysical Observatory, University of Oulu, Tähteläntie 62, FI-99600 Sodankylä
‡ Physics Division, Argonne National Laboratory, Argonne, Illinois 60439, USA
*email address: smajola@uj.ac.za

Low–spin states of 157Dy have been studied using the JUROGAM II array, following the 155Gd (α, 2n) reaction at a beam energy of 25 MeV. The level scheme of 157Dy has been expanded with four new bands. Rotational structures built on the [523]5/2− and [402]3/2+ neutron orbitals constitute new additions to the level scheme as do many of the inter– and intra–band transitions. This manuscript also reports the observation of cross $I^+ \to (I–1)^-$ and $I^- \to (I–1)^+$ E1 dipole transitions inter–linking structures built on the [523]5/2− (band 5) and [402]3/2+ (band 7) neutron orbitals. These interlacing band structures are interpreted as the bands of parity doublets with simplex quantum number $s = –i$ related to possible octupole correlations.

# I. Introduction

Octupole correlations are a manifestation of broken reflection symmetry in the nuclear mean–field. Detailed calculations have been performed to determine the ideal properties of the nuclei which exhibit such a feature. Promising candidates have been studied and octupole correlations have been successfully identified and interpreted using various models [1–34, 88]. The appearance of rotational bands with alternating parity, interconnected by E1 transitions, has been identified as one of the indispensable and common properties of nuclei exhibiting enhanced octupole correlations [1, 10, 16, 35, 36]. The occurrence of large E1 transition probabilities (with collective dipole moments) between the yrast positive– and the negative–parity bands is also expected to create a strong evidence for the presence of octupole deformation, which is generally a combination of an axial quadrupole ($Y_{20}$) and octupole ($Y_{30}$) shapes [1, 49, 51].

In even–even nuclei, octupole correlations involving the ground state and the associated low–lying collective bands are commonly attributed to the interaction between energetically close $\Delta j = \Delta l = 3$ orbitals of opposite parity. In the transitional rare-earth region, the $f_{7/2} - i_{13/2}$ neutron and $h_{11/2} - d_{5/2}$ proton orbitals ($\Delta j = \Delta l = 3$ single–particle nucleons), which interact via octupole coupling, lie close to the Fermi surface for nuclei with N ≈ 88. Thus one can expect a maximum effect here [1, 37]. As a result, over the years, experimental manifestations of static and dynamic octupole deformations have been predicted and successfully confirmed by numerous studies in this region of the nuclear chart [9, 18, 26–28, 37, 38–62].

The odd–mass nuclei can also be a suitable test bed for the presence of octupole deformation at low–spin, particularly if the unpaired valence particle orbiting outside an even–even core occupies an orbit that favors reflection asymmetry. Good examples of reflection asymmetry have been reported for some of the odd–mass nuclei in the N ≈ 90 region, namely the N=91 isotones – $^{153}$Sm and $^{155}$Gd [27, 63]. Due to the structural similarities between the N ≈ 90 isotones in the A ≈ 150–160 mass region [64–70], it is reasonable to expect the presence of octupole correlations in $^{157}$Dy, which is also an N = 91 isotone.

Two previous experiments conducted with large Ge detector arrays used heavy–ion induced reactions $^{150}$Nd($^{12}$C,5n) and $^{124}$Sc($^{36}$S,3n) [71, 72, 97] to study the medium to high–spin states of $^{157}$Dy. These experiments reported rotational bands built on the $h_{9/2}$[521]3/2$^-$, $i_{13/2}$[651]3/2$^+$ and $h_{11/2}$[505]11/2$^-$ neutrons configurations but reported no transitions linking these structures. To this end we have performed in–beam measurements using the JUROGAM II array to study the rotational bands of $^{157}$Dy and to search for evidence of octupole correlations. We report on four new bands as well as many new linking transitions relative to the previously observed structures. The rotational behaviour of the new bands is described in terms of quasi–particle assignments, and the observed alignment properties are analyzed within the framework of the cranked shell model.

## II. Experimental details and analysis

Excited states of $^{157}$Dy were populated using the $^{155}$Gd (α, 2n) reaction at a beam energy of 25 MeV. The $^{155}$Gd target was 0.98 mg/cm$^2$ thick with a purity of 91%. Gamma–rays following the fusion evaporation reaction were detected with the JUROGAM II multi–detector array [73] in JYFL, Jyväskylä, Finland. The spectrometer setup comprised 39 high–purity germanium detectors, all with BGO escape suppression shields; 15 EUROGAM coaxial detectors (Phase 1 [75] and GASP [74] type) and 24 EUROGAM Phase II clover detectors [76]. Approximately 14 x 10$^9$ γ–γ events were unfolded from the data in the off–line analysis and replayed into γ–γ Radware [77, 78] coincidence matrices, which were used to construct the level scheme of $^{157}$Dy.

The spins and parities for the new rotational structures were successfully assigned using the Directional Correlation from Oriented States (DCO ratios/$R_{DCO}$) [79, 80] and linear polarization anisotropy ($A_p$) methods [81]. While $A_p$ is given as used in [82], the DCO matrices were constructed using data from detectors in the rings at 158° and 86°+94°, thus the $R_{DCO}$ for the JUROGAM II array in this work is;

$R_{DCO} = I_{\gamma 1}(at\ 158^o : gated\ on\ \gamma_2\ near\ 90\ ^o)\ /\ I_{\gamma 1}(near\ 90^o : gated\ on\ \gamma_2\ at\ 158^o)$ (1)

When a gate is set on a known stretched quadrupole transition, a $R_{DCO}$ ratio close to 0.5 and 1.0 is expected for pure stretched dipole and quadrupole transitions, respectively. The electric or magnetic character of transitions was inferred from their $A_p$ values. The $A_p$ values yield $A_p > 0$ and $A_p < 0$ for stretched electric and magnetic transitions, respectively.

## III. Experimental results – level scheme

Figures 1 and 2 show partial level schemes of $^{157}$Dy deduced in this work. This includes new bands (colored in red) and rotational structures that have been published in previous in–beam works (colored in black). The new Bands 5, 6, 7 and 8 are labeled in the order in which they were discovered. The level scheme presented in this work focuses on the new bands and structures that they feed into – mainly bands built on the $h_{9/2}[521]3/2^-$ and $i_{13/2}[651]3/2^+$ neutron orbitals. Note that the rotational bands associated with the $h_{11/2}[505]11/2^-$ configuration [71, 72] are not shown in Figures 1 and 2 as no new information relevant to the current study was obtained. Measured properties of γ–rays and rotational levels observed in this work are listed in Table 1.

### A. Bands 1, 2, 3 and 4

The rotational structures labeled bands 1, 2, 3 and 4 have been previously observed up to high–spins [71, 72, 83]. In this work, bands 1 and 2 are also referred to as the ground band while bands 3 and 4 are sometimes referred to as the yrast band. Apart from $R_{DCO}$ and linear polarization measurements confirming the previous spin and parity assignments (see Table 1), there is no new spectroscopic information reported in this study about these bands. Nevertheless, these bands are important as the majority of the band structures reported for the first time here, decay mainly to them (i.e., bands 1, 2, 3 and 4).

## B. Bands 5 and 6

A rotational structure with five in–band transitions, labeled as band 5, has been established in $^{157}$Dy. A γ–ray spectrum showing transitions associated with band 5 is shown in Figure 3 (a). This band decays mainly to both signatures of the ground band (i.e. bands 1 and 2) and is built on the rotational level at 341-keV, which was firmly assigned as a 5/2$^-$ state by Grotdal *et al.,* [84]. Previously, the level at 527-keV in band 5 was also observed and assigned to be either 5/2$^-$ or 7/2$^-$ [84, 85]. However, if the 5/2$^-$ assignment is considered, the inclusion of the newly discovered 365-keV transition that decays to the 9/2$^+$ level in band 3 (see Figure 1) would mean that the 527-keV state would have to decay via a hindered, stretched M2 transition. Similarly, the newly found 229-keV γ–ray, linking the 527-keV level to the 11/2$^+$ state of band 4, would also be an M2, if the 7/2$^-$ assignment were adopted. Given that an M2 transition is unlikely to compete with the in–band transitions, the possibility of both the 5/2$^-$ and 7/2$^-$ assignments to the 527-keV level is eliminated, thus leaving a 9/2$^-$ assignment. The 9/2$^-$ assignment is supported by the $R_{DCO}$ and polarization measurements for the 229 and 466-keV transitions decaying out of the 527-keV level to the 11/2$^+$ and 5/2$^-$ levels of bands 4 and 1, respectively, see Table 1. For instance, the $R_{DCO}$ for the 229-keV γ–ray is close to 0.5 while that of the 466-keV transition is close to 1. This is consistent with them being stretched dipole and quadrupole transitions, respectively, thus supporting our proposed assignment. The $R_{DCO}$ and polarization measurements performed for other transitions (i.e. 269-, 365- and 379-keV γ–rays) decaying out of the 527-keV level are also consistent with it having a 9/2$^-$, see Table 1.

The spin–parity assignments for the remainder of excited levels that form band 5, namely the 827-, 1214-, 1665- and 2154-keV levels have been respectively assigned to 13/2$^-$, 17/2$^-$, 21/2$^-$ and 25/2$^-$. These assignments are also supported by the combined results from the DCO and polarization measurements, carried out for the majority of transition decaying out of these levels.

Band 6 is also new, and is composed of six rotational levels, linked by five stretched E2 transitions. A spectrum showing some of the inter–band and intra–band transitions decaying out of Band 6 is shown in Figure 3 (b). Several transitions decaying out of this band feed into the ground (bands 1 and 2) and yrast bands (bands 3 and 4). The multiple decays out of this band and the $R_{DCO}$ and polarization measurements have made it easy to deduce definitive spins and parities for the first three levels of this band; they are $7/2^-$, $11/2^-$ and $15/2^-$ for the 420-, 672- and 1021-keV levels, respectively. The $R_{DCO}$ measurements for the 586-, 783- and 433-keV transitions, which respectively depopulate the 1021- and 672-keV levels are around 0.5 while the polarization measurements for these transitions yield positive values. This is consistent with them being stretched E1 transitions, thus confirming the proposed spin and parity. Polarization measurements were also carried out for the 1006- and 1159-keV γ–rays, which respectively decay out of the 1441- and 1905-keV levels. Measurements for the 1006- and 1159-keV γ–rays yield $A_p = 0.04(2)$ and $A_p = 0.11(6)$; this is consistent with them being stretched electric transitions.

The $R_{DCO}$ and $A_p$ values were also obtained for some of the in–band members, where possible, and they are consistent with them being stretched E2 transitions, see Table 1.

### C. Bands 7 and 8

A new band, band 7, is based on a level at 235-keV (see Figure 2), which was previously assigned as $3/2^+$ by Grotdal *et al.,* [84]. A gated spectrum showing some of the in–band members of this band is shown in Figure 3 (c). This band is connected to bands 1, 2, 3, 4, 5 and 8 through numerous decay paths, and this has allowed us to determine and confirm the spin and parity assignments for the first four excited states. As a result, the excitation levels at 235-, 435-, 686- and 1025-keV are respectively assigned to have spins and parities of $3/2^+$, $7/2^+$, $11/2^+$ and $15/2^+$. These assignments are corroborated by the combined results of the $R_{DCO}$ and polarization measurements, performed for a number of inter–band transitions, decaying out of this band. Polarization and $R_{DCO}$ measurements have also been performed for all the in–band transitions of this rotational structure, with the exception of the 251-keV γ–ray, and they yield values that are consistent with these transitions being of stretched E2 character, see Table 1. As a result, the excitation levels at 1421-, 1897- and 2393-keV are respectively assigned spins and parities of $19/2^+$, $23/2^+$ and $27/2^+$.

*Figure 1*: (Color on-line) Partial level scheme of $^{157}$Dy deduced from the current work. It shows new rotational structures, bands 5 and 6, which decay predominantly to the ground and yrast bands. New findings from the current work are labeled in red (and marked with asterisk symbols) while previously known bands, deduced from previous in-beam works, are labeled in black. Proposed configurations for each band structure are given below the bands. Quantities within parenthesis are quoted as tentative.

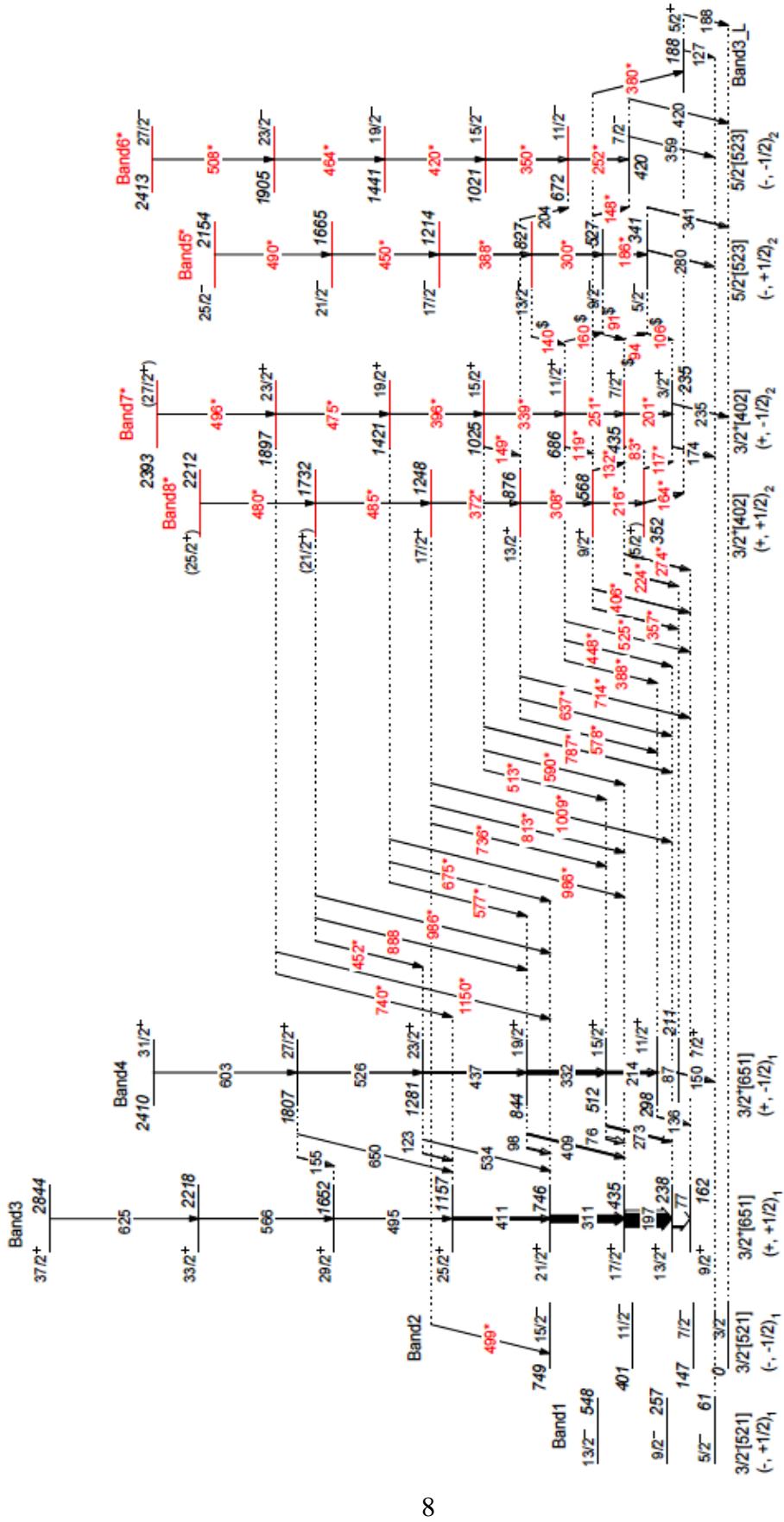

*Figure 2:* (Color on-line) Partial level scheme of $^{157}$Dy deduced from the current work. It shows all the new rotational structures (bansd 5, 6, 7 and 8) built on the yrast and ground bands. New findings from the current work are labeled in red (and marked with asterisk symbols) while previously known bands, deduced from previous in-beam works, are labeled in black.Inter-leaving transitions linking bands 5 and 7 are marked with a dollar symbol ($). Proposed configurations for each band structure are given below the bands. Quantities within parenthesis are quoted as tentative.

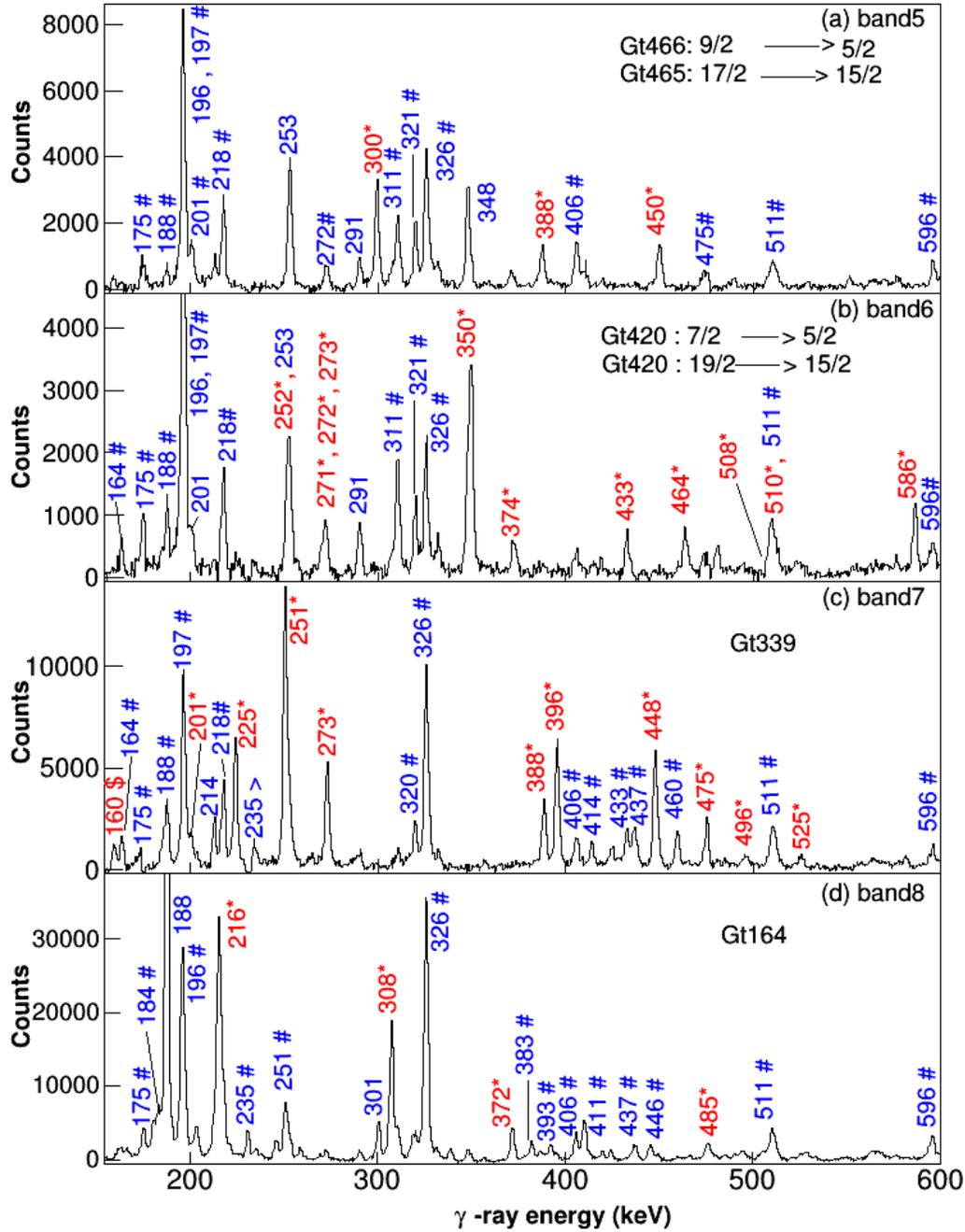

*Figure 3 (Color online): Gated coincidence spectra for bands 5, 6, 7 and 8 are shown in panels (a), (b), (c) and (d), respectively. Note that there is an overlap in the single gates set on doublets in (a) and (b) Transitions corresponding to new structures are labeled in red (and marked with asterisk symbols) while contaminant reaction channels and/or other bands of $^{157}$Dy, not associated with the cascade of interest are denoted by a colored in blue and marked with hash (#) symbols. The transition marked with a dollar symbol ($) in Figure 3 (c) is associated with interleaving E1 transition connecting bands 5 and 7. Previously known transitions in the nucleus and/or decay path of interest are also highlighted in blue but are unmarked.*

Band 8 is a new rotational sequence built on the 352-keV level. This structure decays mainly to the yrast band with a few transitions going to bands 6 and 7. Numerous γ–rays decaying into and out of band 8, fix the spin and parity assignment of the 568- and 876-keV levels to $9/2^+$ and $13/2^+$, respectively. The $R_{DCO}$ and polarization measurements performed for the 380-keV transition linking the 568-keV state to the 188-keV level in band 3_L (an extension of band3) are consistent with it being a stretched electric transition, thus confirming the proposed spin and parity assignment for the 568-keV level. When taking the proposed assignment for the 568-keV level (i.e., $9/2^+$) into account, the spin–parity assignments for the 352-keV level are restricted to $5/2^+$ or $7/2^+$. Assuming that the 352-, 568- and 876-keV levels are members of band 8, connected to each other via stretched E2 transitions leaves the $5/2^+$ as the only possible spin and parity assignment for the 352-keV level. Indeed, the polarization measurement for the 164-keV transition, linking the state in question to the $5/2^+$ level in band 3_L is $A_p = +0.04(1)$ and is consistent with the 164-keV transition being an unstretched M1 transition. The rest of the levels on this band are placed with the assumption that they are band members connected by stretched E2 transitions.

## IV. Discussion

As mentioned above, according to previous studies [71, 72, 83] bands 1 and 2 are built on the ground state of $^{157}$Dy, which have been assigned to the $h_{9/2}[521]3/2^-$ neutron configuration. The configuration of bands 3 and 4 ($i_{13/2}[651]3/2^+$) is also known from previous studies [71, 72]. The rotational behaviour of the new bands is described in terms of quasi–particle assignments and the observed alignment properties are analyzed within the framework of the Cranked Shell Model (CSM) [86, 87]. Figure 4 shows a CSM calculation of Routhians e' or quasi–particle energies of $^{157}$Dy in a rotating frame, deduced using a modified oscillator potential. The labeling of the quasi–neutron states in Figure 4 is given in Table 2. Configurations proposed for the new bands are also given in this table.

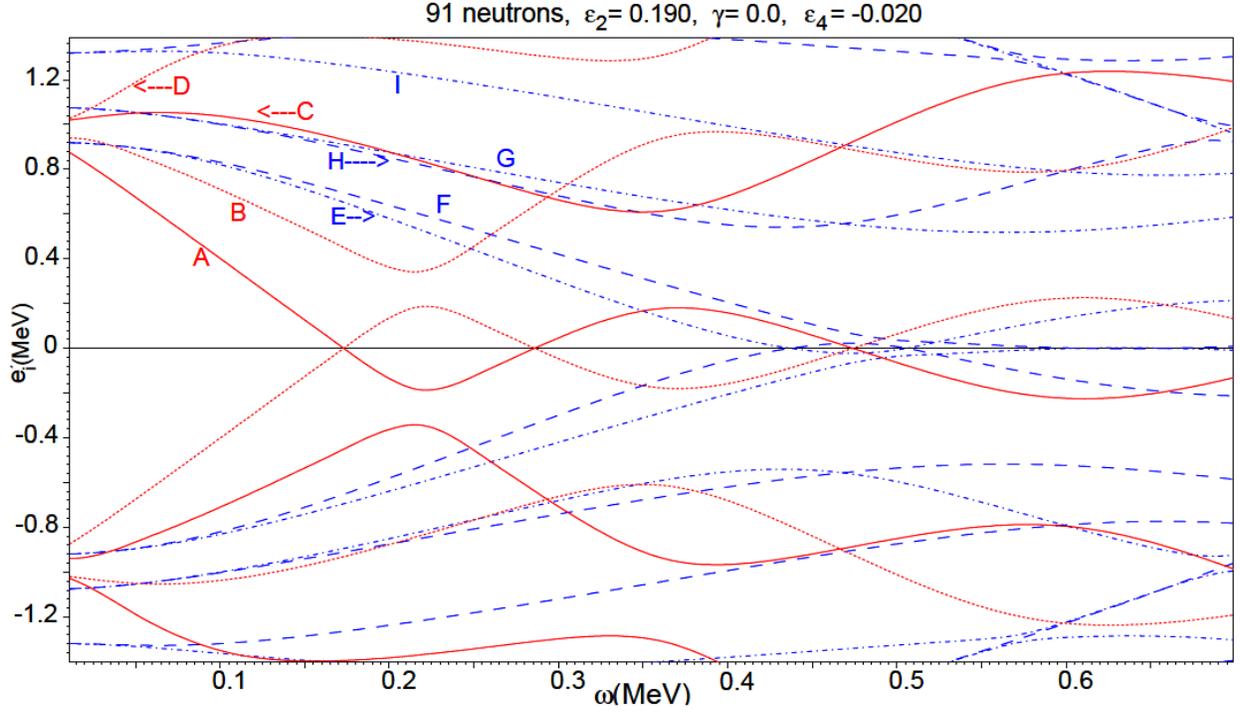

**Figure 4:** *(Color online) Plot of neutron quasi–particle Routhian e', as a function of ℏω, calculated for $^{157}$Dy using the CSM with parameters $\varepsilon_2$ = 0.19, $\varepsilon_4$ = −0.020 and γ = 0°. The solid and dotted lines colored in red represent positive–parity states with signatures α = +1/2 and −1/2, respectively. The dot–dashed and dashed lines colored in blue represent negative–parity states with α = +1/2 and −1/2, respectively. The corresponding quasi–neutron labeling is given in Table 2.*

The trajectories presented in this figure correspond to Routhians of orbitals near the Fermi surface, which are most likely to be actively involved in the formation of the band structures observed experimentally in $^{157}$Dy. The deformation $\varepsilon_2$ = 0.19, used in the calculation, was found by minimization of the total energy for the $\nu i_{13/2}$ neutron band. This deformation is also appropriate for the $h_{9/2}[521]3/2^-$ and $f_{7/2}[523]5/2^-$ neutron bands, but not for the $d_{3/2}[402]3/2^+$ neutron bands. The optimum minima for these were found to be near $\varepsilon_2$ = 0.25. Therefore the Routhian trajectory for the latter are not visible in Figure 4, as they are too high in energy.

**Table 2:** *Labeling of the quasi–neutron states in figure 4 where $(\pi, \alpha)_n$ represent the $n^{th}$ rotational sequence with signature $\alpha$ and parity $\pi$.*

| band label | $(\pi, \alpha)_n$ | Label | Nilsson states |
|---|---|---|---|
| Band1 | $(-,+1/2)_1$ | E | $h_{9/2}[521]3/2^-$ |
| Band2 | $(-,-1/2)_1$ | F | $h_{9/2}[521]3/2^-$ |
| Band3 | $(+,+1/2)_1$ | A | $i_{13/2}[651]3/2^+$ |
| Band4 | $(+,-1/2)_1$ | B | $i_{13/2}[651]3/2^+$ |
| – | $(+,+1/2)_2$ | C | $i_{13/2}[660]1/2^+$ |
| – | $(+,-1/2)_2$ | D | $i_{13/2}[660]1/2^+$ |
| Band5 | $(-,+1/2)_2$ | H | $f_{7/2}[523]5/2^{-*}$ |
| Band6 | $(-,-1/2)_2$ | G | $f_{7/2}[523]5/2^{-*}$ |
| Band7 | $(+,-1/2)_3$ | – | $d_{3/2}[402]3/2^{+\&}$ |
| Band8 | $(+,+1/2)_3$ | – | $d_{3/2}[402]3/2^{+\&}$ |

$^*$ *Strongly mixed with the $h_{9/2}[521]3/2^-$ orbital.*

$^\&$ *Strongly mixed with the $i_{13/2}[651]3/2^+$ orbital.*

## A. Bands 5 and 6

This subsection deals with the quasi–particle excitations responsible for bands 5 and 6. The band head energies of both bands 5 and 6, at 341- and 420-keV, are known from the studies of Grotdal *et al.,* [84] and Klamra *et al.,* [83]. The bands were considered as signature partners associated with the $f_{7/2}[523]5/2^-$ neutron orbital. Indeed, the sequence of levels added in this work, which have allowed us to observe these bands to higher spins, show that these bands share the same moment–of–inertia throughout, as a function of spin, see Figure 5. This feature is also evident in the alignment and Routhian plots, shown in Figures 6 (a) and (b), respectively. In effect, these bands track each other over the observed frequency range. This is consistent with these bands being signature partners built on the same configuration.

In Figure 6 (a), it is apparent that both bands 5 and 6 have an initial alignment close to 1 ℏ. The modest initial alignment of these bands at low frequencies is a strong indication that an $f_{7/2}$ and/or $h_{9/2}$ orbital is involved in the configuration. The CSM calculations are consistent with bands 5 and 6 being signature partners built on the $f_{7/2}[523]5/2^-$ orbital. Firstly, the alignments predicted for both bands 5 and 6 are ≈1.1ℏ, which matches very well with the value experimentally deduced at $\omega = 0.16$-MeV (i.e. ≈ 1.2ℏ), for both bands. According to the quasi–particle Routhian diagram shown in Figure 4, below $\omega = 0.3$ MeV, the slopes corresponding to the $f_{7/2}[523]5/2^-$ configuration (i.e. bands 5 and 6) are predicted to be somewhat parallel to those of structures emanating from the $h_{9/2}[521]3/2^-$ configuration. Indeed, bands 5 and 6 appear to track bands 1 and 2, as predicted in Figure 4. In addition, the relative energy spacing between both structures as well as the order of appearance is also well reproduced, see Figures 4, 5 and 6.

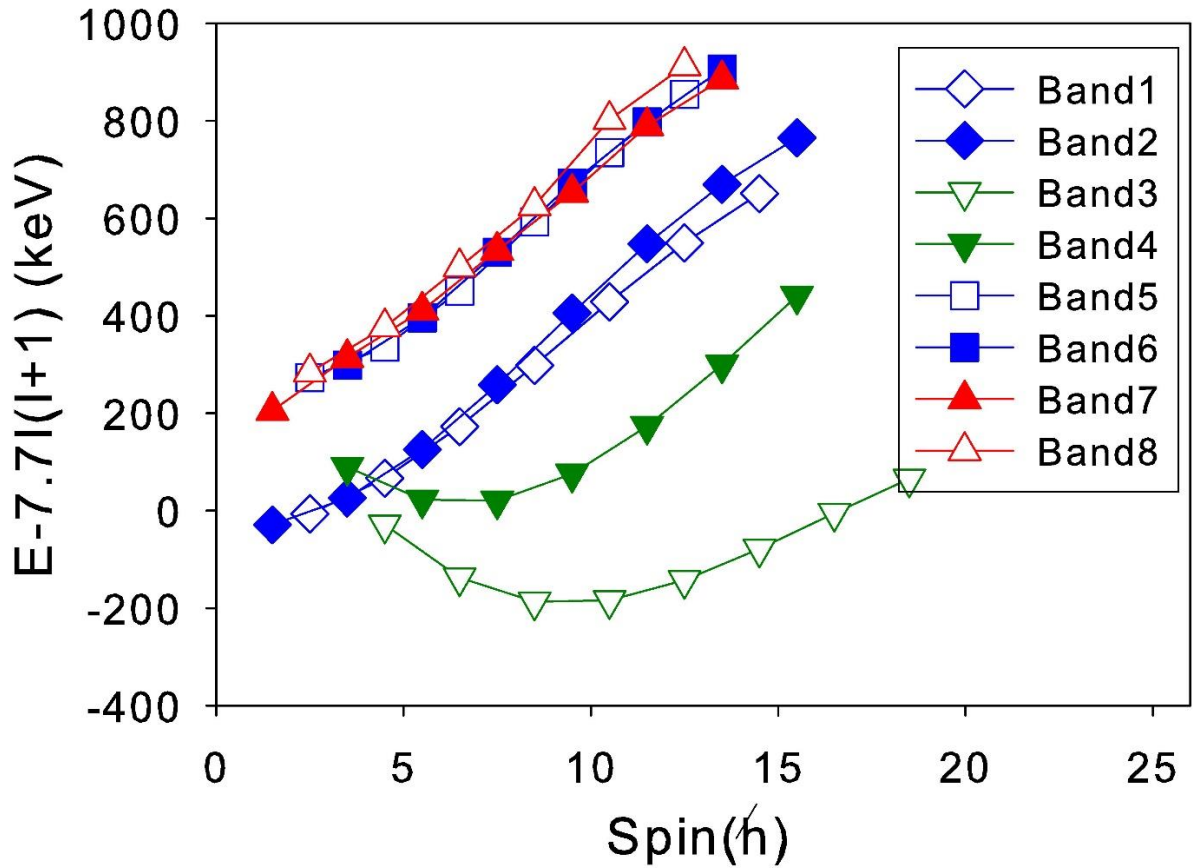

*Figure 5:* (Color online) Plots of excitation energy minus a rigid rotor reference for the bands observed in $^{157}$Dy. Open and closed symbols represent signatures α = +1/2 and α = –1/2, respectively.

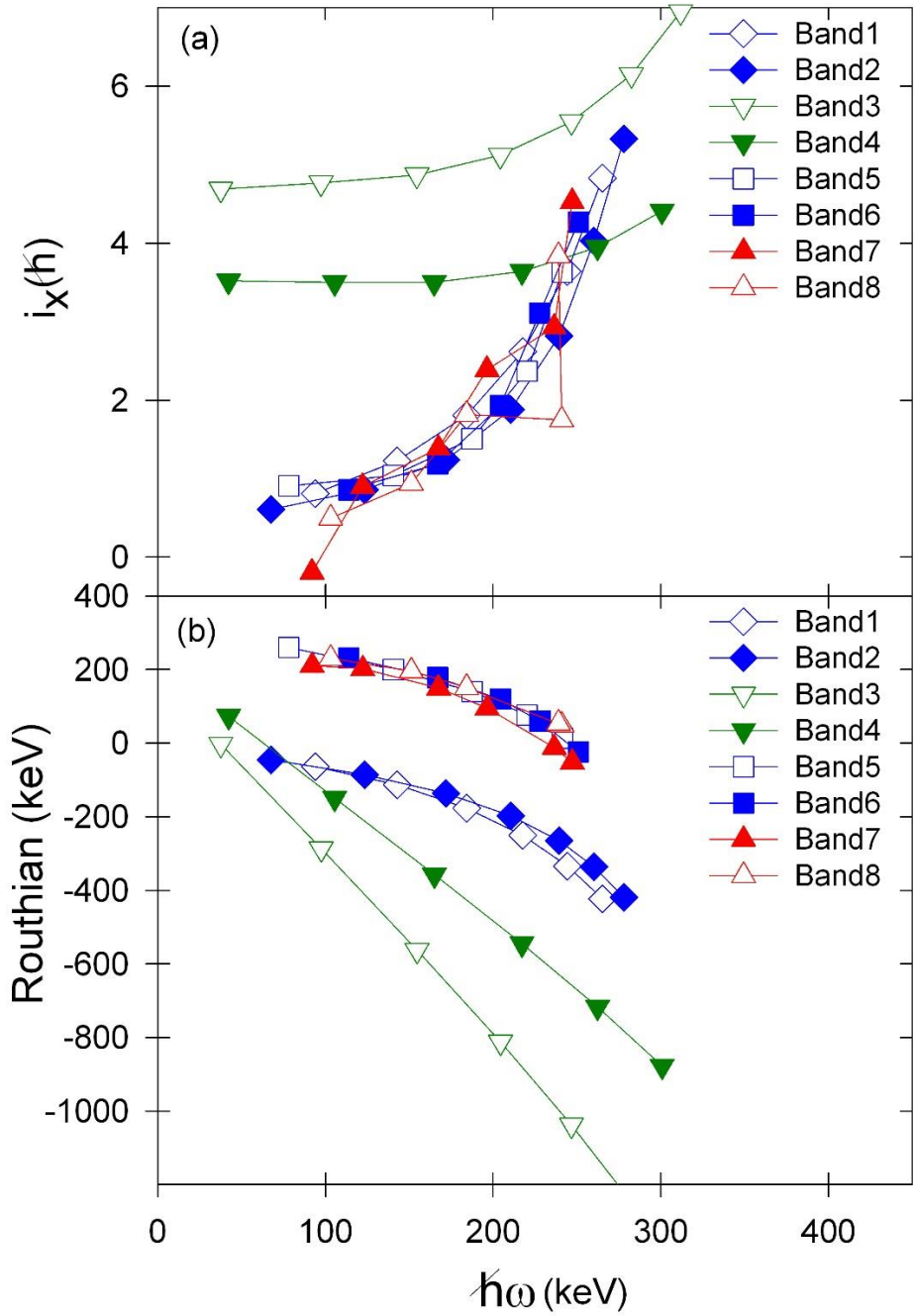

*Figure 6:* (Color online) Plots of the experimental (a) alignment $i_x$ and (b) Routhians e' for the bands in $^{157}$Dy, deduced using Harris parameters of $J_0 = 32\ \hbar^2\ MeV^{-1}$ and $J_1 = 34\ \hbar^4\ MeV^{-3}$. Open and closed symbols represent signatures α = +1/2 and α = −1/2, respectively.

## B. Bands 7 and 8

The interlinking M1 transitions between bands 7 and 8 as well as the manner in which they track one another, both in the alignment and rigid rotor plots, are indicative of them being signature partners. These structures decay mainly to the $i_{13/2}$ bands and are based on a 235-keV level, previously assigned to the $d_{3/2}[402]3/2^+$ neutron configuration [84]. According to [84], this state has admixtures of $i_{13/2}$ neutron orbital, due to the $\Delta N=2$ mixing, caused by the close proximity of the $d_{3/2}[402]3/2^+$ (N=4) and $i_{13/2}[651]3/2^+$ (N=6) Nilsson orbitals at deformation $\varepsilon_2 \approx 0.3$. The majority of the positive–parity bands observed in the odd–even nuclei, near the A ≈ 160 region, are attributed to states associated with the $i_{13/2}$ orbital. In spite of the fact that the bands built on the 235-keV level decay mainly to bands 3 and 4 (the $i_{13/2}$ bands), their structural behavior is somewhat different to that expected for the $i_{13/2}$ bands – in terms of alignment properties, see Figure 6 (a). This is consistent with the fact that the major component of the wave–function of the 235-keV level is dominated by the $d_{3/2}[402]3/2^+$ neutron orbital. Further, the (+, –1/2) signature of the $d_{3/2}[402]3/2^+$ orbital is expected to be the favored one and this is in agreement with what is observed between bands 7 and 8. The apparent upbend observed between $\omega \approx 0.2$ and 0.25 MeV for both bands 7 and 8 in Figure 6 (a) is attributed to the AB crossing following the first alignment of $i_{13/2}$ neutrons [95]. Unlike in the $i_{13/2}$ bands (i.e. bands 3 and 4) where this band crossing is blocked at the cited frequency range, the bands built on $h_{9/2}$ (bands 1 and 2) and $f_{7/2}$ (bands 5 and 6) orbitals are also expected to have similar band crossing frequencies as the $d_{3/2}$ bands (i.e. bands 7 and 8). Indeed, this feature is apparent in the experimental data presented in Figure 6 (a).

## C. Possible enhanced octupole correlations in $^{157}$Dy

Several pronounced E1 transitions are observed between the bands built on the $f_{7/2}[523]5/2^-$ and $i_{13/2}[651]3/2^+$ configurations (the yrast bands), see Figure 1. The occurrence of relatively large E1 transition probabilities between the yrast positive– and negative–parity bands is one of the indispensable properties of nuclei exhibiting the features of reflection asymmetry [1]. However, in the odd–mass rare-earth nuclei, more specifically those with N ≈ 91 such as $^{157}$Dy, it is quite common to find a single–particle negative–parity band decaying predominantly to the yrast positive–parity sequence. In a few instances where such a phenomenon has been observed in the A ≈ 160 mass region, a unique parameter, which accounts for octupole softness has been introduced to help explain this feature [13, 90–93]. The experimental *B(E1)/B(E2)* ratios of out–of–band to in–band transitions for the γ decays from band 6 have been deduced and are listed in Table 3. The weighted average value of the *B(E1)/B(E2)* ratios in band 6 is 0.024(1) ×10$^{-6}$fm$^{-2}$. This value compares very well with the average weighted values of ratios observed in $^{157}$Gd [90], $^{159}$Dy [94], $^{161}$Er [91], $^{163}$Yb [93] and $^{165}$Er [92] where evidence of octupole correlations has been manifested through strong E1 transitions. The relatively enhanced I$^-$ → (I±1)$^+$ decays between bands based on these configurations of alternating parity (i.e [523]5/2$^-$ and [651]3/2$^+$ states), could correspond to the $K = 1^-$ octupole degree of freedom in $^{157}$Dy. This phenomenon may be attributed to the prevalent octupole coupling of valence nucleons occupying the $v(f_{7/2})$ and $v(i_{13/2})$ orbitals, which are close to the Fermi-surface and often form the basis for octupole deformation in the A ≈ 150 − 160 mass region [1]. While these findings might be suggestive that the pronounced I$^-$ → (I ± 1)$^+$ decays between bands based on the $f_{7/2}$[523]5/2$^-$ and $i_{13/2}$[651]3/2$^+$ states, could be due to enhanced octupole correlations, additional experimental quantities, such as *B(E1)s*, would be valuable.

Furthermore, in Figure 2 (and also in Table 1), it is apparent that the rotational sequences of bands 5 and 7 are interlaced with E1 transitions (marked with asterisks in Figure 2), thus forming an alternating parity band. The appearance of rotational bands with alternating parity has been attributed to nuclei having an intrinsic pear shape, which is generated by combining an axial octupole (Y$_{30}$) with an axial quadrupole (Y$_{20}$) shape [1]. Rotational states in octupole deformed nuclei are characterized by eigen values *s* of the simplex operator $\mathscr{S}$ which is related to parity $\mathscr{P}$ and rotation $\mathscr{R}$ operators [23] by:

$$\mathscr{S} = \mathscr{P}\mathscr{R}^{-1} \qquad (2)$$

The parity *p* and spin *I* of a rotational band with simplex *s* are related by:

$$p = se^{-i\pi I} \qquad (3)$$

Consequently, the resulting simplex quantum number value $s$ of an octupole deformed nuclear system with even (integer $I$) number of nucleons is restricted to $s = \pm 1$. In an odd–A nucleus such as $^{157}$Dy, the possible values of the simplex quantum number are $s = \pm i$, such that [1, 23]:

$$s = -i, \qquad I^p = 1/2^-, 3/2^+, 5/2^-, 7/2^+,\ldots \qquad (4)$$
$$s = +i, \qquad I^p = 1/2^+, 3/2^-, 5/2^+, 7/2^-,\ldots \qquad (5)$$

where $I \geq K$. Therefore, in the current context, according to equation (3) bands 5 and 7 form a reflection asymmetric structure with simplex quantum number $s = -i$.

Nuclei which manifest octupole deformation are expected to have large dipole moments ($D_0$), thus giving rise to fast E1 transitions, which are expected to compete with intra–band E2 transitions. The experimental $B(E1)/B(E2)$ ratios of the out–of–band to the in–band transitions for the γ decays from the $s = -i$ band have also been deduced in this work and they are listed in Table 3. The weighted average for this ratio is 0.057(28) ×10$^{-6}$fm$^{-2}$ for the $s = -i$ candidate in $^{157}$Dy. The value is an order of magnitude smaller compared to the $s = -i$ bands of $^{143}$Ba and $^{145}$Ba [61]. This could imply a modest value for the intrinsic dipole moment. Alternatively, this could also be caused by the larger quadrupole deformation of $^{157}$Dy. Indeed, it is known that the prolate deformation of $^{157}$Dy is much larger than that of Ba isotopes, which are situated in the island of octupole deformation. In effect, the sudden increase in quadrupole deformation at and beyond N ≈ 90 may trigger a change in the intrinsic shell structure, which makes the observation of reflection asymmetric shapes rather difficult. In addition, octupole correlations can also be diluted by the intermediate $h_{9/2}$ neutron orbital which comes close to the Fermi level and lies between the $\Delta j = 3$ neutron orbitals – further making it more difficult to confirm the reflection asymmetric configurations [27].

Furthermore, information on the nature of the octopule deformation exhibited by a nucleus can be inferred from the experimental energy displacement $\delta E(I)$, as given by [23]:

$$\delta E(I) = E(I^-) - \frac{1}{2}[E(I+1)^+ + E(I-1)^+] \qquad (6)$$

In the case of $^{157}$Dy, this is plotted for the $s = -i$ band as shown in Figure 7. Here we compare it to similar structures that have been used to confirm the existence of strong octupole correlations in the rare-earth ($Z \approx 56$, $N \approx 88$) and in the actinide ($Z \approx 90$, $N \approx 134$) regions, namely $^{143}$Ba [61] and $^{146}$Nd [44]), $^{220}$Ra [88] and $^{223}$Th [89], respectively.

Clearly, the $s = -i$ structure compares well with simplex structures found in the islands of stable octupole correlations in the ($Z \approx 56$, $N \approx 88$) and ($Z \approx 90$, $N \approx 134$) regions. Furthermore, the energy displacement $\delta E(I)$ is expected to approach 0 for structures exhibiting permanent (or stable) octupole deformation and this feature is evident in the $s = -i$ structure in $^{157}$Dy. In effect, this structure is within 5% nearer to $\delta E(I) = 0$ and is the closest to the stable octupole deformation limit (throughout as function of spin $I$) compared to all other simplex structures shown in Figure 7.

Another approach that can be used to examine the nature of the octupole deformation is through the rotational frequency ratio $\omega^-(I)/\omega^+(I)$ given by [23]:

$$\frac{\omega^-(I)}{\omega^+(I)} = 2\frac{E(I+1)^- - E(I-1)^-}{E(I+2)^+ - E(I-2)^+} \qquad (7)$$

This ratio is expected to approach 1 when nuclei evolve to stable octupole deformation. In order to gain more insight about the octupole correlations in $^{157}$Dy, we plot the $\omega^-(I)/\omega^+(I)$ ratio in Figure 8 and compare it to nuclei with octupole enhanced octupole correlations. Here it is clear that the $\omega^-(I)/\omega^+(I)$ ratio for $^{157}$Dy is nearer to 1 (within 5% for $I > 5$) whereas it approaches the stable octupole deformation limit at relatively high–spin (spin $I \approx 9$) for simplex structures found in the well–established octupole nuclei (Z≈56 N≈88 and Z≈90, N≈134) regions. These results are in accordance with what is expected for a nucleus exhibiting stable octupole deformation.

While the aforementioned bands manifest features that are suggestive of enhanced octupole correlations in $^{157}$Dy, it is worth noting that these features cannot be considered as unambiguous proof. It should be noted that there are cases whereby interleaving bands, assigned with the same K-value, could be explained without invoking octupole deformation [2, 17, 20, 96]. Therefore, it is in general, desirable to understand better the features of bands with enhanced octupole correlations. Indeed, additional theoretical and experimental spectroscopic studies (such as lifetime measurements) are thus needed in order to obtain further insight on the existence of enhanced octupole correlation in $^{157}$Dy.

Octupole correlations in nuclei are understood to be triggered by the octupole–octupole interaction, which arises between valence nucleons that occupy pairs of orbitals with $\Delta l = 3$ [1, 36]. Band 5 is associated with the $f_{7/2}$ orbital, which mixes strongly with the $h_{9/2}$ orbital. Therefore, if the parity-doublet bands in $^{157}$Dy are indeed due to the enhanced octupole correlations, the alternating parity observed between bands 5 and 7 may be attributed to valence nucleons occupying the $h_{9/2}$ and $d_{3/2}$ orbitals, which are likely to be involved in the formation of these bands, such that $\Delta j = 9/2 - 3/2 = 3$.

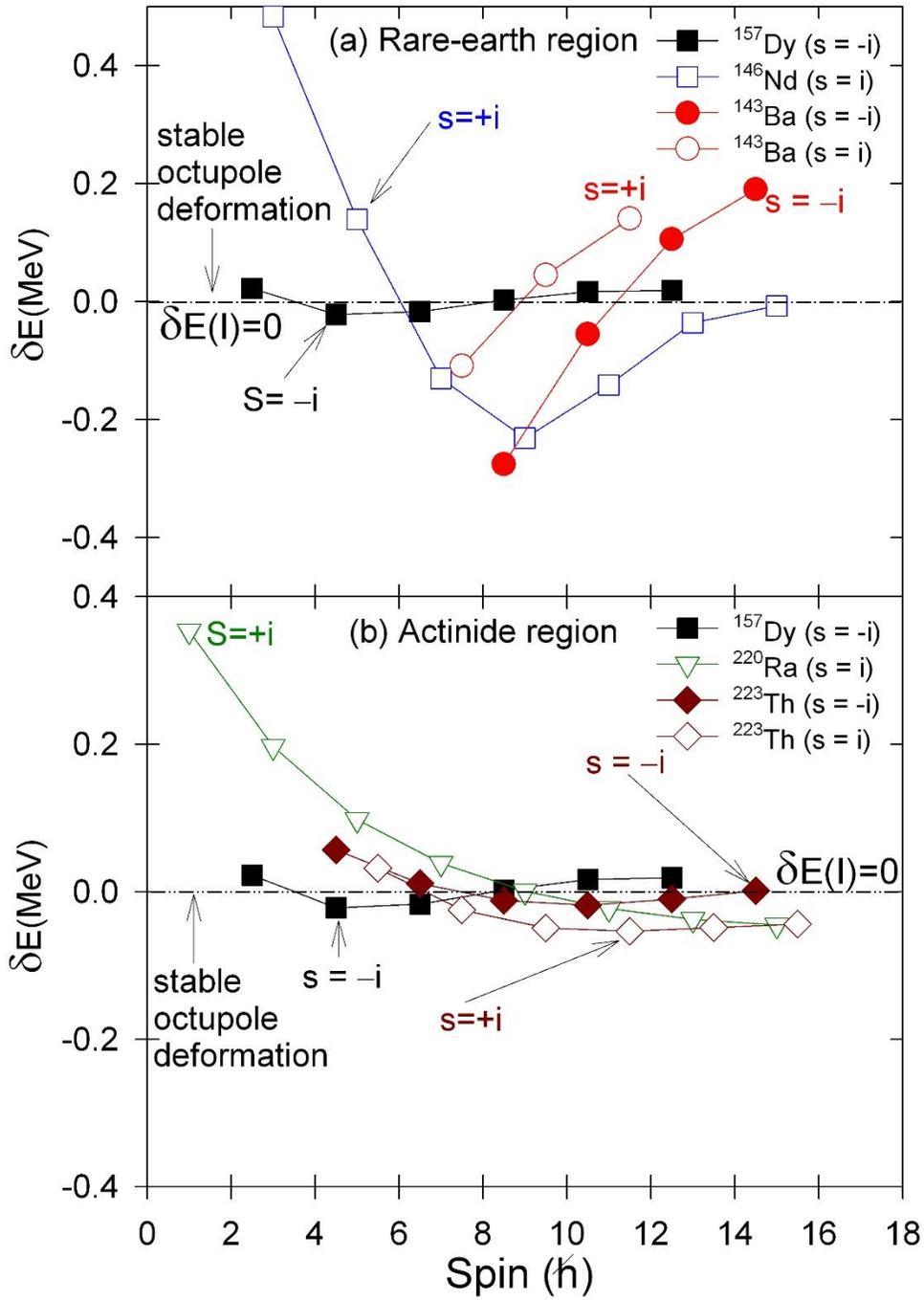

*Figure 7: (Color online) Plot of the energy displacement δE(I) of bands 5 and 7 forming a reflection asymmetric structure (with simplex s = −i) in $^{157}$Dy. This structure is compared to similar structures identified in nuclei with considerable octupole deformation in the (a) rare-earth region ($^{143}$Ba [61] and $^{146}$Nd [44]) and (b) actinide region ($^{220}$Ra [88] and $^{223}$Th [89]). The energy displacement δE(I) is expected to approach 0 for structures exhibiting permanent octupole deformation. Open and closed symbols correspond to s = +i and s = −i simplex structures, respectively.*

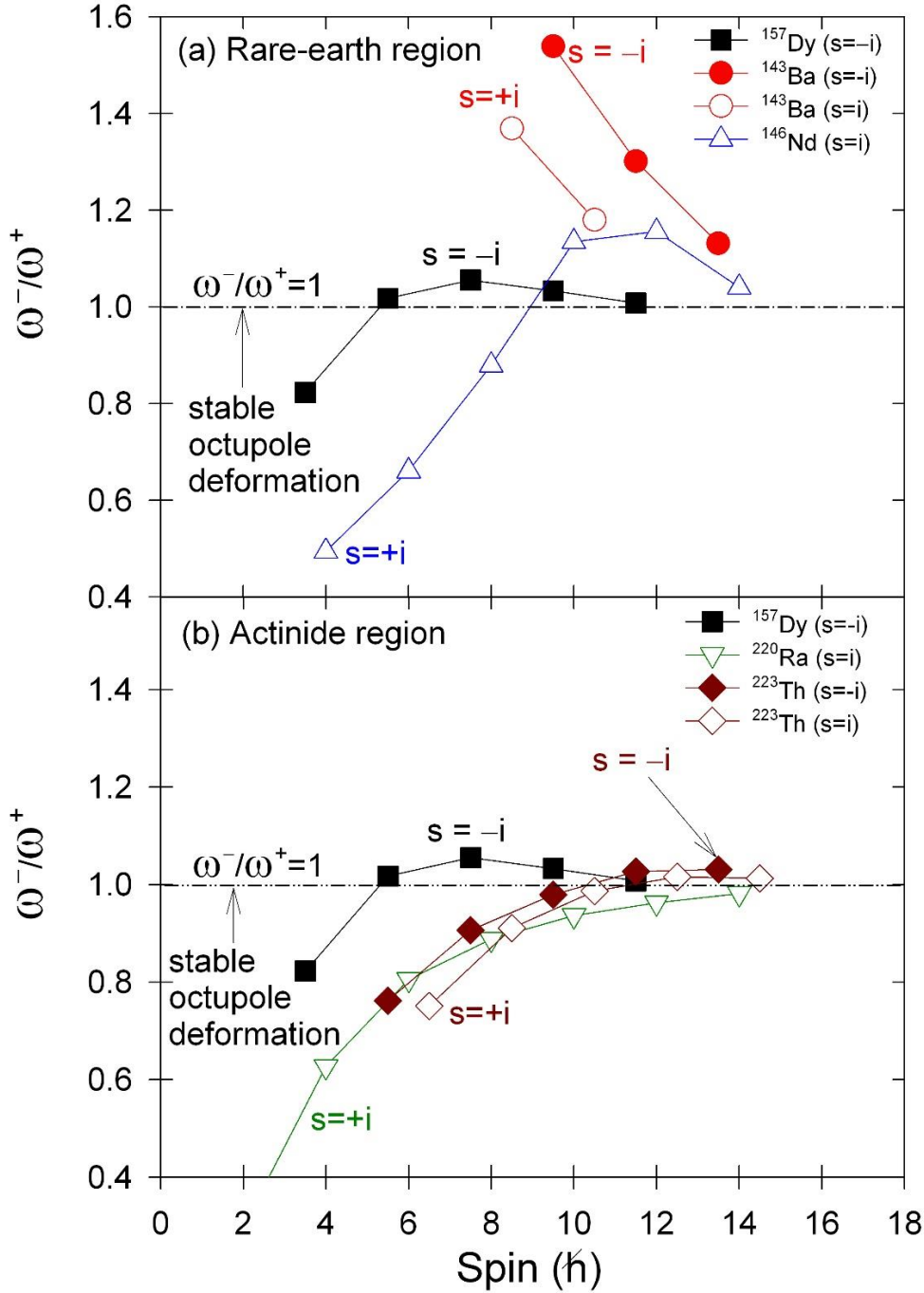

*Figure 8:* (Color online) Rotational frequency ratio $\omega^-(I)/\omega^+(I)$ of bands 5 and 7 forming a reflection asymmetric structure (with simplex $s = -i$) in $^{157}$Dy. This structure is compared to similar structures identified in nuclei with considerable octupole deformation in the (a) rare-earth region ($^{143}$Ba [61] and $^{146}$Nd [44]) and (b) in the actinide region ($^{220}$Ra [88] and $^{223}$Th [89]). This ratio is expected to approach 1 for structures exhibiting stable octupole deformation. Open and closed symbols correspond to $s = +i$ and $s = -i$ simplex structures, respectively.

*Table 3*: The B(E1)/B(E2) ratios of out–of–band to in–band transitions for the γ decays from the s = –i candidate (band 5 and 7) and band 6 (to the yrast bands) in $^{157}$Dy.

| $I_i^\pi \to I_f^\pi$ | E1(γ–energy in keV) | B(E1)/B(E2), ($\times 10^{-6}$fm$^{-2}$) |
|---|---|---|
| s = –i | | |
| **Bands 5 and 7** | | |
| $7/2^+ \to 5/2^-$ | 94 | 0.039(10) |
| $9/2^- \to 7/2^+$ | 91 | 0.027(3) |
| $11/2^+ \to 9/2^-$ | 160 | 0.064(8) |
| $13/2^- \to 11/2^+$ | 140 | 0.096(7) |
| **Band6** | | |
| $11/2^- \to 9/2^+$ | 510 | 0.018(2) |
| $11/2^- \to 13/2^+$ | 433 | 0.018(1) |
| $15/2^- \to 13/2^+$ | 783 | 0.007(1) |
| $15/2^- \to 17/2^+$ | 586 | 0.012(1) |
| $19/2^- \to 17/2^+$ | 1006 | 0.022(1) |
| $19/2^- \to 21/2^+$ | 695 | 0.049(2) |
| $23/2^- \to 21/2^+$ | 1159 | 0.030(2) |
| $23/2^- \to 25/2^+$ | 748 | 0.042(2) |

## V. Summary and conclusion

An experiment populating the low–spin states of $^{157}$Dy has been performed to investigate the possibility of octupole correlations in this nucleus. The level scheme of $^{157}$Dy has been expanded, at low–spin, with four new bands. Bands 5, 6, 7, and 8 constitute new additions to the level scheme of $^{157}$Dy, as do many of the intra– and inter–band transitions. Candidates for signature partner bands built on both the $f_{7/2}$[523]5/2$^-$ and $d_{3/2}$[402]3/2$^+$ neutron orbitals have been identified for the first time.

This experiment also reports the observation of cross $I^+ \to (I–1)^-$ and $I^- \to (I–1)^+$ E1 dipole transitions interlinking structures built on the [523]5/2$^-$ and [402]3/2$^+$ orbitals. The energy displacement $\delta E(I)$ and rotational frequency ratio $\omega^-(I)/\omega^+(I)$ of the resulting parity doublet, formed by interlacing bands 5 and 7, with simplex quantum number $s = –i$ are approximately equal to 0 and 1, respectively. While these bands manifest features that are necessary and essential for a nucleus exhibiting the breaking of the intrinsic reflection symmetry, it is worth noting that they cannot be considered as unambiguous proof. Therefore, future experimental and theoretical investigative studies are essential in order to positively identify whether the alternating-parity rotational bands, observed in $^{157}$Dy, are indeed associated with enhanced octupole correlation.

## VI. ACKNOWLEDGEMENTS

This work is supported by the South African National Research Foundation under grants (No. 96829, 109711, 93531, 109134, 116666 and 106012). Support for L.B. and P.E.G. was provided by the Natural Sciences and Engineering Research Council of Canada. JYFL research is supported by the Academy of Finland under the Finnish Centre of Excellence Programme 2006–2011, Contract No. 213503. This work was also partially supported by the National Research, Development and Innovation Fund of Hungary, financed under the K18 funding scheme with project no. K128947, as well as by the European Regional Development Fund (Contract No. GINOP–2.3.3–15–2016–00034). Support for M.R. was provided by the U.S. National Science Foundation under Grant Nos. PHY-1502092 (USNA) and PHY–0754674 (FSU). The authors also acknowledge the support of GAMMAPOOL for the loan of the JUROGAM detectors.

***Table 1:*** *Experimentally determined properties for the nucleus* $^{157}Dy$. *This includes excitation energy levels* $E_x$ *(in keV),* $\gamma$–*ray energies* $E_\gamma$ *(in keV),* $\gamma$–*ray intensities* $I_\gamma$, *assigned multipolarities (Mult), spins (for the initial and final states* $I_i^\pi$ *and* $I_f^\pi$, *respectively), level in the band of destination Band#, DCO ratios (*$R_{DCO}$*) and polarization measurements* $(A_p)$. *The* $R_{DCO}$ *ratios were deduced by gating on stretched E2 transitions with the exception of those marked by an asterisk, which were deduced by gating on a pure stretched dipole. Quantities within parenthesis are quoted as tentative while empty entries refer to information that could not be obtained.*

| $E_x(keV)$ | $E_\gamma(keV)$[1] | $I_\gamma$ | Mult | $I_i^\pi$ | $I_f^\pi$ | Band# | $R_{DCO}$ | $A_p$ |
|---|---|---|---|---|---|---|---|---|
| *Band1* | | | | | | | | |
| 61 | 61.0 | 21.6(16) | M1/E2 | 5/2⁻ | 3/2⁻ | Band2 | | |
| 257 | 109.9 | 18.7(6) | M1/E2 | 9/2⁻ | 7/2⁻ | Band2 | | |
| 257 | 196.4 | 19.8(7) | E2 | 9/2⁻ | 5/2⁻ | Band1 | | 0.13(3) |
| 548 | 147.1 | 9.3(3) | M1/E2 | 13/2⁻ | 11/2⁻ | Band2 | | |
| 548 | 290.6 | 22.7(7) | E2 | 13/2⁻ | 9/2⁻ | Band1 | 0.85(16) | 0.07(1) |
| 920 | 171.0 | 2.25(7) | M1/E2 | 17/2⁻ | 15/2⁻ | Band2 | | |
| 920 | 372.3 | 14.1(4) | E2 | 17/2⁻ | 13/2⁻ | Band1 | 0.99(10) | 0.09(2) |
| 1358 | 184.5 | 0.85(3) | M1/E2 | 21/2⁻ | 19/2⁻ | Band2 | | |
| 1358 | 438.2 | 6.42(19) | E2 | 21/2⁻ | 17/2⁻ | Band1 | 1.04(12) | 0.08(2) |
| 1849 | 490.7 | 2.31(7) | E2 | 25/2⁻ | 21/2⁻ | Band1 | 1.11(30) | |
| 2381 | 532.2 | 0.57(2) | E2 | 29/2⁻ | 25/2⁻ | Band1 | 1.40(49) | |
| *Band2* | | | | | | | | |
| 147 | 86.5 | 44.7(14) | M1/E2 | 7/2⁻ | 5/2⁻ | Band1 | | |
| 147 | 147.5 | 15.6(6) | E2 | 7/2⁻ | 3/2⁻ | Band2 | | |
| 401 | 143.6 | 14.8(5) | M1/E2 | 11/2⁻ | 9/2⁻ | Band1 | | |
| 401 | 253.4 | 20.1(6) | E2 | 11/2⁻ | 7/2⁻ | Band2 | | |
| 749 | 201.2 | 5.51(17) | M1/E2 | 15/2⁻ | 13/2⁻ | Band1 | | –0.08(8) |
| 749 | 348.3 | 15.8(5) | E2 | 15/2⁻ | 11/2⁻ | Band2 | 0.89(16) | 0.09(2) |
| 1174 | 253.5 | 1.69(6) | M1/E2 | 19/2⁻ | 17/2⁻ | Band1 | | –0.08(2) |
| 1174 | 424.5 | 7.37(22) | E2 | 19/2⁻ | 15/2⁻ | Band2 | 0.93(16) | 0.11(3) |
| 1655 | 296.5 | 0.61(3) | M1/E2 | 23/2⁻ | 21/2⁻ | Band1 | | |
| 1655 | 481.3 | 2.96(9) | E2 | 23/2⁻ | 19/2⁻ | Band2 | 0.99(17) | 0.09(3) |
| 2177 | 522.2 | 0.89(3) | E2 | 27/2⁻ | 23/2⁻ | Band2 | 0.91(17) | 0.09(5) |
| 2735 | 557.6 | 0.20(1) | E2 | 31/2⁻ | 27/2⁻ | Band2 | 1.03(6) | |
| *Band3* | | | | | | | | |
| 238 | 76.5 | ≈100 | E2 | 13/2⁺ | 9/2⁺ | Band3 | | |
| 435 | 196.9 | 98(3) | E2 | 17/2⁺ | 13/2⁺ | Band3 | | 0.13(3) |
| 746 | 311.0 | 52.1(16) | E2 | 21/2⁺ | 17/2⁺ | Band3 | 0.83(8) | 0.10(2) |

| $E_x(keV)$ | $E_\gamma(keV)$[1] | $I_\gamma$ | Mult | $I_i^\pi$ | $I_f^\pi$ | Band# | $R_{DCO}$ | $A_p$ |
|---|---|---|---|---|---|---|---|---|
| | | | | *Band3 continued* | | | | |
| 1157 | 410.8 | 16.6(5) | E2 | 25/2$^+$ | 21/2$^+$ | Band3 | 0.92(10) | 0.08(2) |
| 1652 | 495.1 | 4.10(13) | E2 | 29/2$^+$ | 25/2$^+$ | Band3 | 1.19(16) | 0.06(2) |
| 2218 | 566.4 | 0.89(3) | E2 | 33/2$^+$ | 29/2$^+$ | Band3 | 0.86(16) | 0.04(5) |
| 2844 | 625.5 | 0.25(1) | E2 | 37/2$^+$ | 33/2$^+$ | Band3 | | |
| | | | | *Band 3_L* | | | | |
| 188 | 127.2 | 0.05(2) | E1 | 5/2$^+$ | 5/2$^-$ | Band1 | | |
| 188 | 188.2 | 0.5(2) | E1 | 5/2$^+$ | 3/2$^-$ | Band2 | | 0.38(15) |
| | | | | *Band4* | | | | |
| 211 | 150.0 | 3.35(10) | E1 | 7/2$^+$ | 5/2$^-$ | Band1 | | |
| 298 | 86.8 | 7.16(10) | E2 | 11/2$^+$ | 7/2$^+$ | Band4 | | |
| 298 | 136.2 | 1.58(16) | M1/E2 | 11/2$^+$ | 9/2$^+$ | Band3 | | |
| 512 | 76.5 | 5.65(20) | M1/E2 | 15/2$^+$ | 17/2$^+$ | Band3 | | |
| 512 | 213.7 | 16.9(6) | E2 | 15/2$^+$ | 11/2$^+$ | Band4 | | 0.11(3) |
| 512 | 273.5 | 10.2(4) | M1/E2 | 15/2$^+$ | 13/2$^+$ | Band3 | 0.66(8) | –0.03(2) |
| 844 | 97.8 | 2.58(8) | M1/E2 | 19/2$^+$ | 21/2$^+$ | Band3 | | |
| 844 | 332.4 | 24.0(7) | E2 | 19/2$^+$ | 15/2$^+$ | Band4 | 0.89(8) | 0.09(2) |
| 844 | 408.8 | 9.1(3) | M1/E2 | 19/2$^+$ | 17/2$^+$ | Band3 | | |
| 1281 | 123.5 | 0.70(3) | M1/E2 | 23/2$^+$ | 25/2$^+$ | Band3 | | |
| 1281 | 436.7 | 11.4(3) | E2 | 23/2$^+$ | 19/2$^+$ | Band4 | 1.12(10) | 0.08(1) |
| 1281 | 534.5 | 1.70(6) | M1/E2 | 23/2$^+$ | 21/2$^+$ | Band3 | | |
| 1807 | 154.9 | 0.18(2) | M1/E2 | 27/2$^+$ | 29/2$^+$ | Band3 | | |
| 1807 | 526.5 | 3.04(9) | E2 | 27/2$^+$ | 23/2$^+$ | Band4 | 1.12(8) | 0.71(4) |
| 1807 | 649.9 | 0.52(3) | M1/E2 | 27/2$^+$ | 25/2$^+$ | Band3 | | |
| 2410 | 603.2 | 0.84(4) | E2 | 31/2$^+$ | 27/2$^+$ | Band4 | | |
| | | | | *Band5* | | | | |
| 341 | 106.5 | 0.13(3) | E1 | 5/2$^-$ | 3/2$^+$ | Band7 | | |
| 341 | 152.8 | 0.17(2) | E1 | 5/2$^-$ | 5/2$^+$ | Band 3_L | | |
| 341 | 280.0 | 0.61(6) | M1/E2 | 5/2$^-$ | 5/2$^-$ | Band1 | 0.71(7) | |
| 341 | 341.0 | 0.41(5) | M1/E2 | 5/2$^-$ | 3/2$^-$ | Band2 | | |
| 527 | 91.5 | 0.29(3) | E1 | 9/2$^-$ | 7/2$^+$ | Band7 | | |
| 527 | 106.6 | 0.47(4) | M1/E2 | 9/2$^-$ | 7/2$^-$ | Band6 | | |
| 527 | 125.7 | 1.12(5) | M1/E2 | 9/2$^-$ | 11/2$^-$ | Band2 | | |
| 527 | 185.6 | 0.99(5) | E2 | 9/2$^-$ | 5/2$^-$ | Band5 | | |
| 527 | 228.8 | 1.20(8) | E1 | 9/2$^-$ | 11/2$^+$ | Band4 | 0.52(31) | 0.03(2) |
| 527 | 269.5 | 2.21(10) | M1/E2 | 9/2$^-$ | 9/2$^-$ | Band1 | 0.87(22) | –0.27(11) |
| 527 | 365.0 | 0.83(6) | E1 | 9/2$^-$ | 9/2$^+$ | Band3 | 1.12(4) | –0.07(2) |
| 527 | 379.5 | 0.81(5) | M1/E2 | 9/2$^-$ | 7/2$^-$ | Band2 | 0.61(7) | |
| 527 | 465.7 | 1.34(7) | E2 | 9/2$^-$ | 5/2$^-$ | Band1 | 1.23(3) | |

| $E_x(keV)$ | $E_\gamma(keV)^1$ | $I_\gamma$ | Mult | $I_i^\pi$ | $I_f^\pi$ | Band# | $R_{DCO}$ | $A_p$ |
|---|---|---|---|---|---|---|---|---|
| | | | | Band5 continued | | | | |
| 827 | 140.5 | 0.56(3) | E1 | 13/2⁻ | 11/2⁺ | Band7 | | |
| 827 | 154.9 | 0.55(4) | M1/E2 | 13/2⁻ | 11/2⁻ | Band6 | | |
| 827 | 278.5 | 0.68(4) | M1/E2 | 13/2⁻ | 13/2⁻ | Band1 | | |
| 827 | 299.9 | 3.93(13) | E2 | 13/2⁻ | 9/2⁻ | Band5 | 0.83(23) | 0.09(4) |
| 827 | 314.9 | 0.63(2) | E1 | 13/2⁻ | 15/2⁺ | Band4 | 0.42(18) | |
| 827 | 425.6 | 2.15(8) | M1/E2 | 13/2⁻ | 11/2⁻ | Band2 | 0.47(27) | –0.02(2) |
| 827 | 569.5 | 2.30(9) | E2 | 13/2– | 9/2⁻ | Band1 | 1.11(6) | 0.07(7) |
| 1214 | 294.3 | 0.031(2) | M1/E2 | 17/2– | 17/2⁻ | Band1 | | |
| 1214 | 388.0 | 3.19(10) | E2 | 17/2– | 13/2⁻ | Band5 | 1.15(3) | 0.11(2) |
| 1214 | 465.3 | 0.90(3) | M1/E2 | 17/2– | 15/2⁻ | Band1 | | |
| 1214 | 666.5 | 0.52(3) | E2 | 17/2– | 13/2⁻ | Band1 | | |
| 1214 | 779.3 | 0.64(4) | E1 | 17/2⁻ | 17/2⁺ | Band3 | | |
| 1665 | 450.2 | 1.50(5) | E2 | 21/2⁻ | 17/2⁻ | Band5 | 1.23(28) | 0.04(2) |
| 1665 | 491.1 | 0.17(2) | M1/E2 | 21/2⁻ | 19/2⁻ | Band2 | 0.93(3) | 0.07(2) |
| 2154 | 489.5 | 0.35(2) | E2 | 25/2⁻ | 21/2⁻ | Band5 | | |
| | | | | Band6 | | | | |
| 420 | 162.7 | 2.10(11) | M1/E2 | 7/2⁻ | 9/2⁻ | Band1 | 0.54(17) | |
| 420 | 209.0 | 2.00(19) | E1 | 7/2⁻ | 7/2⁺ | Band4 | 0.95(8) | |
| 420 | 258.4 | 2.16(18) | E1 | 7/2⁻ | 9/2⁺ | Band3 | | |
| 420 | 272.5 | 4.81(22) | M1/E2 | 7/2⁻ | 7/2⁻ | Band2 | | |
| 420 | 359.1 | 0.72(11) | M1/E2 | 7/2⁻ | 5/2⁻ | Band1 | | |
| 420 | 420.0 | 1.02(13) | E2 | 7/2⁻ | 3/2⁻ | Band2 | 2.19(10)* | |
| 672 | 123.6 | 0.40(3) | M1/E2 | 11/2⁻ | 13/2⁻ | Band1 | | |
| 672 | 251.6 | 1.41(7) | E2 | 11/2⁻ | 7/2⁻ | Band6 | | |
| 672 | 270.7 | 1.13(6) | M1/E2 | 11/2⁻ | 11/2⁻ | Band2 | | |
| 672 | 373.8 | 1.12(9) | E1 | 11/2⁻ | 11/2⁺ | Band4 | | |
| 672 | 414.5 | 1.28(7) | M1/E2 | 11/2⁻ | 9/2⁻ | Band1 | | |
| 672 | 433.3 | 2.63(12) | E1 | 11/2⁻ | 13/2⁺ | Band3 | | 0.12(1) |
| 672 | 510.0 | 4.4(3) | E1 | 11/2⁻ | 9/2⁺ | Band3 | | |
| 672 | 524.5 | 0.76(6) | E2 | 11/2⁻ | 7/2⁻ | Band2 | | |
| 1021 | 272.1 | 0.61(4) | M1/E2 | 15/2⁻ | 15/2⁻ | Band2 | | |
| 1021 | 349.7 | 2.58(9) | E2 | 15/2⁻ | 11/2⁻ | Band6 | | 0.23(8) |
| 1021 | 473.3 | 0.59(3) | M1/E2 | 15/2⁻ | 13/2⁻ | Band1 | | |
| 1021 | 586.1 | 1.56(7) | E1 | 15/2⁻ | 17/2⁺ | Band3 | 0.54(31) | 0.07(2) |
| 1021 | 620.4 | 0.77(4) | E2 | 15/2⁻ | 11/2⁻ | Band2 | | |
| 1021 | 783.0 | 2.18(10) | E1 | 15/2⁻ | 13/2⁺ | Band3 | 0.74(27) | 0.12(8) |
| 1441 | 419.6 | 1.43(5) | E2 | 19/2⁻ | 15/2⁻ | Band6 | 2.20(11)* | 0.07(1) |

| $E_x(keV)$ | $E_\gamma(keV)^1$ | $I_\gamma$ | Mult | $I_i^\pi$ | $I_f^\pi$ | Band# | $R_{DCO}$ | $A_p$ |
|---|---|---|---|---|---|---|---|---|
| \multicolumn{9}{c}{***Band6 continued***} ||||||||
| 1441 | 694.7 | 2.32(8) | E1 | 19/2⁻ | 21/2⁺ | Band3 | | |
| 1441 | 1005.7 | 3.23(11) | E1 | 19/2⁻ | 17/2⁺ | Band3 | | 0.04(2) |
| 1905 | 463.9 | 0.52(2) | E2 | 23/2⁻ | 19/2⁻ | Band6 | 0.89(4) | 0.18(8) |
| 1905 | 747.8 | 0.55(2) | E1 | 23/2⁻ | 25/2⁺ | Band3 | | |
| 1905 | 1158.9 | 1.48(5) | E1 | 23/2– | 21/2⁺ | Band3 | | 0.11(6) |
| 2413 | 507.8 | 0.83(3) | E2 | 27/2– | 23/2⁻ | Band6 | | |
| 2413 | 1255.6 | 0.61(2) | E1 | 27/2– | 25/2⁺ | Band3 | | |
| \multicolumn{9}{c}{***Band7***} ||||||||
| 235 | 234.6 | 2.6(4) | E1 | 3/2⁺ | 3/2⁻ | Band2 | | –0.18(2) |
| 235 | 173.6 | 2.0(3) | E1 | 3/2⁺ | 5/2⁻ | Band1 | | |
| 435 | 83.3 | 0.37(3) | M1/E2 | 7/2⁺ | (5/2⁺) | Band8 | | |
| 435 | 94.4 | 0.12(3) | E1 | 7/2⁺ | 5/2⁻ | Band5 | | |
| 435 | 200.8 | 0.91(5) | E2 | 7/2⁺ | 3/2⁺ | Band8 | | 0.06(1) |
| 435 | 224.4 | 3.19(23) | M1/E2 | 7/2⁺ | 7/2⁺ | Band4 | 0.80(18) | |
| 435 | 273.7 | 2.12(17) | M1/E2 | 7/2⁺ | 9/2⁺ | Band3 | | |
| 686 | 118.5 | 0.34(3) | M1/E2 | 11/2⁺ | 9/2⁺ | Band10 | | |
| 686 | 159.5 | 0.46(4) | E1 | 11/2⁺ | 9/2⁻ | Band5 | | |
| 686 | 251.0 | 3.13(11) | E2 | 11/2⁺ | 7/2⁺ | Band7 | | |
| 686 | 388.3 | 0.97(6) | M1/E2 | 11/2⁺ | 11/2⁺ | Band4 | | 0.09(7) |
| 686 | 448.0 | 1.84(9) | M1/E2 | 11/2⁺ | 13/2⁺ | Band3 | | |
| 686 | 524.5 | 0.41(6) | M1/E2 | 11/2⁺ | 9/2⁺ | Band3 | | |
| 1025 | 149.4 | 0.55(3) | M1/E2 | 15/2⁺ | 13/2⁺ | Band8 | | |
| 1025 | 339.0 | 1.98(7) | E2 | 15/2⁺ | 11/2⁺ | Band7 | 1.12(19) | 0.17(2) |
| 1025 | 513.4 | 0.47(16) | M1/E2 | 15/2⁺ | 15/2⁺ | Band4 | | |
| 1025 | 590.1 | 1.39(6) | M1/E2 | 15/2⁺ | 17/2⁺ | Band3 | | |
| 1025 | 787.0 | 2.89(12) | M1/E2 | 15/2⁺ | 13/2⁺ | Band3 | | –0.04(1) |
| 1421 | 395.8 | 1.83(6) | E2 | 19/2⁺ | 15/2⁺ | Band7 | 0.95(6) | 0.06(6) |
| 1421 | 577.5 | 1.14(5) | M1/E2 | 19/2⁺ | 19/2⁺ | Band4 | 0.68(31) | |
| 1421 | 674.9 | 0.53(3) | M1/E2 | 19/2⁺ | 21/2⁺ | Band3 | 0.61(25) | |
| 1421 | 985.9 | 1.39(6) | M1/E2 | 19/2⁺ | 17/2⁺ | Band3 | | |
| 1897 | 475.5 | 1.25(5) | E2 | 23/2⁺ | 19/2⁺ | Band7 | | 0.28(8) |
| 1897 | 739.5 | 0.48(2) | M1/E2 | 23/2⁺ | 25/2⁺ | Band3 | | |
| 1897 | 1150.6 | 0.55(3) | M1/E2 | 23/2⁺ | 21/2⁺ | Band3 | | |
| 2393 | 496.2 | 0.64(3) | E2 | (27/2⁺) | 23/2⁺ | Band7 | 0.883(103) | |
| \multicolumn{9}{c}{***Band8***} ||||||||
| 352 | 117.4 | 0.19(3) | M1/E2 | (5/2⁺) | 3/2⁺ | Band7 | | |
| 352 | 163.9 | 2.41(8) | M1/E2 | (5/2⁺) | 5/2⁺ | Band 3_L | | 0.04(1) |
| 568 | 132.0 | 0.35(3) | M1/E2 | 9/2⁺ | 7/2⁺ | Band7 | | |

| $E_x(keV)$ | $E_\gamma(keV)^1$ | $I_\gamma$ | Mult | $I_i^\pi$ | $I_f^\pi$ | Band# | $R_{DCO}$ | $A_p$ |
|---|---|---|---|---|---|---|---|---|
| | | | | *Band8 continued* | | | | |
| 568 | 147.6 | 1.48(8) | E1 | $9/2^+$ | $7/2^-$ | Band6 | | |
| 568 | 215.6 | 2.01(8) | E2 | $9/2^+$ | $(5/2^+)$ | Band8 | | |
| 568 | 356.6 | 0.48(6) | M1/E2 | $9/2^+$ | $7/2^+$ | Band4 | | |
| 568 | 380.0 | 0.40(3) | E2 | $9/2^+$ | $5/2^+$ | Band 3_L | 2.49(15)* | 0.11(2) |
| 568 | 405.9 | 5.08(24) | M1/E2 | $9/2^+$ | $9/2^+$ | Band3 | | |
| 876 | 203.9 | 0.76(5) | E1 | $13/2^+$ | $11/2^-$ | Band6 | | |
| 876 | 307.8 | 4.0(1) | E2 | $13/2^+$ | $9/2^+$ | Band8 | | 0.06(6) |
| 876 | 577.7 | 1.02(8) | M1/E2 | $13/2^+$ | $11/2^+$ | Band4 | | |
| 876 | 637.5 | 2.31(11) | M1/E2 | $13/2^+$ | $13/2^+$ | Band3 | | 0.09(7) |
| 876 | 713.9 | 1.58(16) | E2 | $13/2^+$ | $9/2^+$ | Band3 | | |
| 1248 | 372.0 | 2.61(9) | E2 | $17/2^+$ | $13/2^+$ | Band8 | | |
| 1248 | 498.6 | 0.46(3) | E1 | $17/2^+$ | $15/2^-$ | Band2 | | |
| 1248 | 736.4 | 2.23(9) | M1/E2 | $17/2^+$ | $15/2^+$ | Band4 | | |
| 1248 | 812.5 | 1.95(9) | M1/E2 | $17/2^+$ | $17/2^+$ | Band3 | | |
| 1248 | 1009.5 | 1.58(16) | E2 | $17/2^+$ | $13/2^+$ | Band3 | | |
| 1732 | 451.5 | 0.45(2) | M1/E2 | $(21/2^+)$ | $23/2^+$ | Band4 | | |
| 1732 | 484.5 | 0.75(3) | E2 | $(21/2^+)$ | $17/2^+$ | Band8 | | 0.06(9) |
| 1732 | 888.2 | 0.91(4) | M1/E2 | $(21/2^+)$ | $19/2^+$ | Band4 | | |
| 1732 | 986.2 | 1.54(20) | M1/E2 | $(21/2^+)$ | $21/2^+$ | Band3 | | |
| 2212 | 480.1 | 0.49(3) | E2 | $(25/2^+)$ | $21/2^+$ | Band8 | | |

[1]Typical uncertainty of 0.3 keV and up to 0.5 keV for weak transitions ($I_\gamma<1$).